\documentclass[3p]{elsarticle}
\usepackage[dvipsnames]{xcolor} 

\usepackage{textcomp}	
\usepackage{bm}
\usepackage{gensymb}
\usepackage{multirow}
\usepackage{booktabs}
\usepackage{tabularx}
\usepackage{todonotes}
\usepackage[normalem]{ulem}
\usepackage{xcolor}
\usepackage{diagbox}
\usepackage{amsmath}
\usepackage{lineno,hyperref}
\usepackage{soul}
\usepackage{todonotes}

\modulolinenumbers[1]
\journal{Journal of \LaTeX\ Templates}

\usepackage{framed}
\usepackage{multicol}
\usepackage{nomencl}

\newcommand{\bs}{\boldsymbol}

\makenomenclature
\renewcommand*\nompreamble{\begin{multicols}{2}}
\renewcommand*\nompostamble{\end{multicols}}

\newcommand{\mysize}{0.65} 
\newcommand{\mmmysize}{0.4} 

\definecolor{greengrg}{RGB}{60,150,80}

\usepackage{booktabs,xcolor,colortbl}
\usepackage{xparse}
\ExplSyntaxOn
\NewDocumentCommand{\longdash}{ O{2} }
 {
  --\prg_replicate:nn { #1 - 1 } { \negthinspace -- }
 }
\ExplSyntaxOff

\usepackage{pifont}

\begin{document}
\begin{frontmatter}

\title{Experimental investigation and numerical modelling of density-driven segregation in an annular shear cell}  

\author[address1]{Monica Tirapelle}
\author[address1]{Andrea C. Santomaso\corref{secondcorrespondingauthor}}
\cortext[secondcorrespondingauthor]{Principal corresponding author. \\ $Email$ $address:$ andrea.santomaso@unipd.it}
\author[address2]{Patrick Richard}
\author[address2]{Riccardo Artoni}

\address[address1]{APTLab - Advanced Particle Technology Laboratory, Department of Industrial Engineering, University of Padova, Italy}
\address[address2]{MAST-GPEM, University Gustave Eiffel, IFSTTAR, F-44344 Bouguenais, France}

\newpage
\begin{abstract}
Granular materials segregate spontaneously due to differences in particle size, shape, density and flow behaviour. In this paper we experimentally investigate density-difference-driven segregation for a range of density ratios and a range of heavy particle concentrations. The experiments are conducted in an annular shear cell with rotating bumpy bottom that yields an exponential shear profile. The cell is initially filled with a layer of light particles and an upper layer of heavier grains and, on top, a load provides confinement. The segregation process is filmed through the transparent side-wall with a camera, and the evolution of particle concentration in space and time is evaluated by means of post-processing image analysis.
We also propose a continuum-approach to model density-driven segregation. We use a segregation-diffusion transport equation, constitutive relations for effective viscosity and friction coefficient, and a segregation velocity analogous to the Stokes' law. The model, which is validated by comparison with experimental findings, can successfully predict density-driven segregation at different density ratios and volumetric fraction. 
\end{abstract}

\end{frontmatter}

\section{Introduction}
Segregation of cohesionless particulate materials is one of the major problems that plagues all industries involved with the handling and processing of bulk solids \cite{Williams1976, Standish1985}. Because of segregation, mixture quality deteriorates, and production costs and wastes rise.

Segregation occurs when particles differ in size, shape, density, flowability, surface roughness or resilience \cite{Williams1976,Rosato1986,tang2004}. Despite particle size is known to be the most relevant factor determining segregation, the effect of large density differences may also be of importance for some industries \cite{Williams1976}. Because of density differences, the lighter particles may drift towards the top of the sheared layer, whereas the denser particles sink to the bottom \cite{Savage1988, Alonso1991}.

Even if the majority of the published studies deals with size segregation, a number of studies have been conducted with a view to understand density-driven segregation in dense granular flows using both experiments \cite{Drahun1983, Alonso1991, Hill1999, Shi2007, Hill2008, Sanfratello2009, Xiao2016} and discrete element (DEM) simulations \cite{Tripathi2011, Tripathi2013, Tunuguntla2014, Fan2015, Duan2019a}. Furthermore, numerous theoretical models have been reported in the literature \cite{Tripathi2013, Tunuguntla2014, Gray2015, Xiao2016, Duan2019a}. In the development of such theories two different approaches have been used. The first one is based on the Kinetic Theory of Granular Flow (KTGF), whereas the second one is based on a hydrodynamic balance of forces. 
In the former case each species is characterized by its own granular temperature. The granular temperature is a kinetic energy associated with velocity variances, and it drives segregation \cite{Fan2015, Yoon2006}. Despite it has been shown that the kinetic theory predicts well segregation in the case of low or moderate solids fractions, it breaks down quantitatively in the case of densely packed flows, which are characterized not only by collisions but also by multi-body long-lasting contacts with neighbouring particles \cite{Pouliquen2002, Rajchenbach2004, Fan2015}. 
On the other hand, models based on hydrodynamic forces (drag and buoyancy) are able to successfully reproduce gravity-driven segregation \cite{Khakhar1997, Tripathi2013} although they have phenomenological bases.

Density-driven segregation has been studied in a variety of systems, among which there are fluidized beds \cite{Zeilstra2008}, vertically vibrated cylinders \cite{Jain2013b}, rotating cylinders \cite{Hajra2011a, Khakhar2003, Sanfratello2009}, inclined chutes \cite{Tripathi2011, Tunuguntla2014, Gray2015}, split-bottom cells \cite{Hill2008} and shear cells \cite{Savage1984, Wildman2008, May2010, Artoni2018}. 

Here, experiments are performed in an annular shear cell with a rotating bumpy bottom and an overloaded top bumpy wall. The geometry is similar to that used in Savage and Sayed \cite{Savage1984}, and Wildman et al. \cite{Wildman2008}; furthermore it has been used by May et al. \cite{May2010} to study shear-driven size segregation and by Artoni et al. \cite{Artoni2018} to asses stresses, shear localization and wall friction of confined dense granular flows. 
This geometry has many advantages: it allows the generation of a continuous granular flow and the assessment of well-defined initial and final states. Moreover, in this type of systems, the velocity profile decays exponentially with depth (from the rotating shearing bottom to the top lid) and hence, the shear rate is not uniform. To the author's knowledge, nobody has studied density-driven segregation in such a system before.

The cell is filled with a binary mixture of spherical particles, all having approximately the same size (6 mm diameter) but different densities. The particle bed is initially stratified: a layer of heavier spheres is poured on a layer of less denser grains. A top bumpy wall free to move vertically, with some masses loaded on it, sits above them. The percolation process is recorded at the side-wall through a camera and the concentration profiles is evaluated with post-processing image analysis. The experiments are conducted for a range of density ratios and heavy particle concentrations. 

In this paper, we also propose a mathematical model for describing density-driven segregation of dense granular flows. The model is similar to the segregation model previously proposed by Tripathi and Khakhar \cite{Tripathi2013}: as they did, we derived a relation for the segregation velocity based on a modified version of the Stokes' law. However, partly because of the relatively simple flow configuration, our formulation is characterized by a significant reduction in complexity: we introduced neither the effective temperature nor a coupling with a rheological model for describing the flow field. 
The latter is made possible by assuming a velocity profile known a priori. The aim was to reduce the uncertainties related to a rheological model and to focus mainly on definition and validation of the segregation law. The velocity profile is expected to depend on the local pressure profile.
This segregation velocity is implemented within a transport equation for the species concentration, namely a segregation-diffusion equation. The system is closed with constitutive relations for both granular viscosity and friction coefficient taken from literature \cite{Jop2006}.
The one-dimensional model is numerically solved for the same density ratios and initial particle concentrations that has been investigated experimentally. 

To assess the validity and the accuracy of the model, the numerical results are compared with the experimental findings. In particular, the root-mean-square deviation (RMSD) between the measured and predicted evolution of the heavy particle concentration, which evolves in space and time, has been estimated and used to obtain the optimal model parameters, which are correlated to particle and contact properties. It is shown that the model is able to predict density-driven segregation of dense granular flow, also when grains are subjected to exponential shear rate profiles, provided that, the degree of exponential decay appearing in the velocity profile is correctly estimated. We then found that density segregation is very sensitive to the shear localization features and that, at low load, a transverse friction coefficient may induce a three dimensional flow pattern.

The paper is organized as follows. Next section ($\S$\ref{SECT: Exp. method}) presents the experimental set-up and the measurements method; whereas section $\S$\ref{SECT: Num. model} deals with the mathematical model. The experimental and numerical results are presented in sections $\S$\ref{SECT: Result_exp} and $\S$\ref{SECT: Result_num}, respectively. We will discuss in detail the effect of the exponential decay in section $\S$\ref{SECT: discussion}.
The conclusions of the work are presented in section $\S$\ref{SECT: conclusions}.

\section{Experimental method} \label{SECT: Exp. method}
\subsection{Experimental set-up and operative conditions} \label{SUBSECT: Apparatus}
Experiments were conducted with $\approx6$ mm ($D_p$) diameter spherical particles confined in an annular geometry made of two coaxial cylindrical poly-methyl-meth-acrylate (PMMA) walls. The smallest cylinder had outer diameter, $D_{in}$, equal to 90 mm, whereas the largest cylinder had inner diameter, $D_{out}$, of 190 mm. The annular region was therefore 50 mm thick.  

\begin{figure}
	\centering \includegraphics[width=\mmmysize\textwidth]{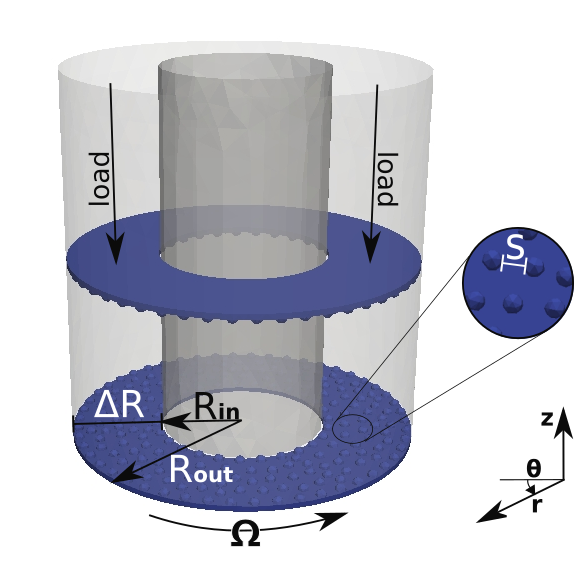}
	\caption{Sketch of the annular shear cell and the coordinate system. Particles are filled in 	the annular region with thickness equal to $\Delta R$ in between the two horizontal bumpy walls. The bottom wall rotates at a \textcolor{black}{rotational speed} $\Omega$; whereas on the top lid additional load is applied.}
	\label{FIG: geometry}
	\end{figure}

The horizontal top and bottom boundaries were made of poly-lactic acid (PLA) and printed with a 3D printer. To increase friction between walls and grains and to reduce sliding, these horizontal walls were designed with hemispheres of diameter $d_{em}=D_p$ placed on a continuous random triangulation with a mean distance of $s=2D_p$ between their edges. 
The bottom surface was fixed on a rotating plate, whereas the top surface was free to move vertically, but could not rotate. On the top surface, some masses could also be loaded in order to vary the compressive force applied on the particle bed. A sketch of the experimental apparatus is reported in Figure \ref{FIG: geometry}. 
For a more detailed description of the set-up, see Artoni et al. \cite{Artoni2018}.

All the experiments that are reported in this paper were performed with the bottom plate rotating at a constant \textcolor{black}{rotational speed} of $\Omega=23.44$ rpm; furthermore a weight $M_w$ of $1.093$ kg (top bumpy wall plus steel masses) was always applied on the grains.

\begin{figure*}
	\centering
	\begin{center} \includegraphics[width=0.75\textwidth]{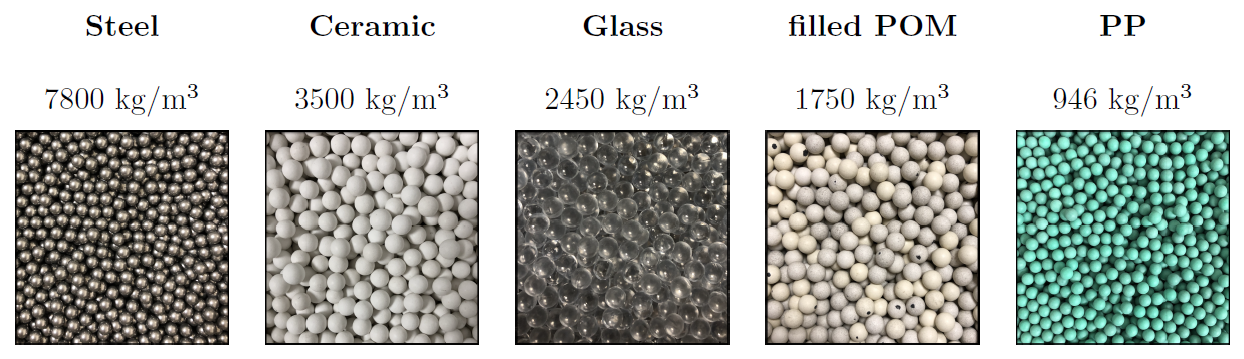}
	\end{center}
	\caption{Photographs of the materials and their corresponding densities. From the left to the right: Steel, Ceramic, Glass, filled Polyoxymethylene and Polypropylene. 
\textcolor{black}{Their particle diameters are: $6.0\pm0.025$ mm for steel, $6.0\pm 1.0$ mm for ceramic (however we measured $d_{10}=5.6$ mm and $d_{90}=6.3$ mm), $6.0\pm0.3$ mm for glass, $5.9\pm0.1$ mm for filled Polyoxymethylene and $6.0\pm0.05$ mm for Polypropylene.}	}
	\label{FIG: 5 materials}
	\end{figure*}

At the start of each experiment, the cell was filled with two layers of spherical particles: a lower layer of lighter particles and an upper layer of more massive particles. 
The position of the interface between the two layers therefore depended on the initial overall volumetric fractions (or concentrations) of the two species, which we denote, in reference to the heavy particles, as $c_{h,0}$ (the dual concentration of the light particles is $c_{l,0}= 1-c_{h,0}$).
The total bed height, $H$, was equal to 90 mm, namely 15 particle diameters.
To increase the number of possible combinations of density ratio, five different-density spherical particles were employed. They were made of Stainless Steel, Ceramic, Glass, filled Polyoxymethylene and Polypropylene (see Fig. \ref{FIG: 5 materials}). 
All the combinations of density ratio, which is defined as $\delta_R \equiv \frac{\rho_h}{\rho_l}$, and volume fraction for which segregation has been evaluated, are listed in Table \ref{TAB: Exp. test}. For the tested combinations, the amount of grains filled in the cell, $M_g$, is reported in kg. We will see later that mass differences (the mass varied in the range $\left[1.62, 8.64\right]$ kg) have a huge impact on the localization of the shear band, and by implication on segregation.

\textcolor{black}{Preliminary experiments highlighted that the tests were fairly repeatable. In the framework of a qualitative comparison with a segregation model, we present in the following only one repetition for each tested condition.}

\begin{table} \footnotesize \centering
	\begin{tabular}{lllrccccc}
	 &  &  &  & \multicolumn{5}{c}{\textbf{$M_g$ [kg]}} \\ \cline{5-9} 
	\textbf{ID} & \textbf{Heavy p.} & \textbf{Light p.} & \textbf{$\delta_R$} &  10\%  &  25\% &  50\% &  75\%  & 90\%  \\ \hline
	S-PP & Steel & Polypropylene & 8.25 & \longdash[4] & \longdash[4] & 5.25 & 7.30 & 8.54 \\
	S-POM & Steel & Filled POM & 4.46 & 2.83 & 3.92 & 5.73 & 7.55 & 8.64 \\
	C-PP & Ceramic & Polypropylene & 3.70 & \longdash[4] & \longdash[4] & 2.67 & 3.43 & 3.89 \\
	G-PP & Glass & Polypropylene & 2.59 & \longdash[4] & \longdash[4] & 2.04 & 2.49 & 2.76 \\
	POM-PP & Filled POM & Polypropylene & 1.85 & \longdash[4] & \longdash[4] & 1.62 & \longdash[4] & \longdash[4] \\
	G-POM & Glass & Filled POM & 1.40 & \longdash[4] & \longdash[4] & 2.52 & \longdash[4] & \longdash[4]\\	\hline
	\end{tabular}
	\caption{Details of the experimental conditions. The columns report, from left to right, the combination ID, the heavy component, the light component, their density ratio $\delta_R$ and $c_{h,0}$. For the tested combinations, the total mass of grains $M_g$ filled in the cell is reported in kg.} \label{TAB: Exp. test}
	\end{table}

\subsection{Concentration measurements at the wall} \label{SUBSECT: Concentration estimation} 
The segregation process was filmed through the transparent outer cylindrical wall by a commercial camera (GoPro Hero 4 black) placed besides the cell. The camera was running at 0.5 or 1.5 fps, depending on the  investigated density ratio. In the case of $\delta_R=1.40,$ $\delta_R=1.85$ and $\delta_R=2.59$, the segregation process was such slow that 0.5 fps was small enough to capture the main features of segregation. At the two sides of the camera, two homogeneous light emitting diode (LED) lamps were placed for lighting the system. Once taken, the images were transferred from the camera to the computer.

\begin{figure*}
\centering
\includegraphics[width=\mysize\textwidth]{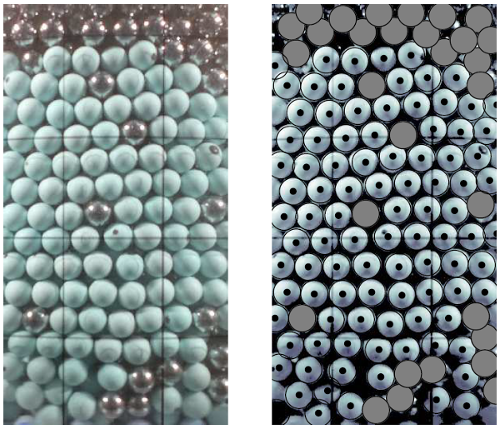} 
\caption{On the left: example of a cropped and undistorted picture. On the right: draw of the detected particles distinguished for density.}
\label{FIG: particle count}
\end{figure*}

We measured the concentration profile at the sidewall by post-processing image analysis. The images, which were slightly distorted because taken with a wide-angle lens, were firstly undistorted, cropped, and successively converted from the RGB colour space to Grey. The particles were detected thanks to the HoughCircles tool \cite{opencv} that gave as output the coordinates of the centres and the diameters of the particles. 
To discern heavy and light particles, the cropped RGB images were also converted in the HSV or LAB colour-space and subjected to threshold segmentation. These images were used as mask for discriminating between the two material types. After post-processing, we ended up with knowing position and type of each particle in time. An example of processed image is shown Fig. \ref{FIG: particle count}. On the left, the frame is reported after distortion correction, whereas on the right the detected particles, differentiated by type, are drawn. 

The window for the investigation spanned the entire height of the particle bed ($H=15D_p$) and was 9 $D_p$ wide (small enough to reduce the effect of the wall curvature). To estimate how the concentration evolves in space and time, the investigation window was subdivided into 15 discrete horizontal layers, each of which was one particle diameter high. The concentration of heavy particles in a given layer was calculated as the number of heavy particles over the total number of particles having their centre within that layer. To reduce the noise in the concentration profiles, the moving average technique was applied over time. 

To summarize, we first determined the type and position of each particle by means of post-processing image analysis. Then, the temporal evolution of the heavy particle concentration through the depth of the cell was reconstructed.

\section{Density segregation model} \label{SECT: Num. model}
As it was shown, in the annular shear cell the flow can be approximated as unidirectional in the azimuthal direction $\theta$, whereas it varies only in the vertical $z$-direction: $\bs{v}=\left(0,v_{\theta}(z),0\right)$. In this section, a continuum model for describing density-driven segregation is developed.

\subsection{Conservation laws}\label{SUBSECT: conservation law}
For describing the change in local constituent concentration due to segregation, each species has to satisfy a transport equation. Here it is assumed a diffusive-segregating transport mechanism, so the volume concentration $c_i$, which is defined as the ratio between the local volume fraction of specie $i$ and the total solid volume fraction ($c_i=\phi_i/\phi_s$), is given by:
\begin{equation}
\frac{\partial c_i}{\partial t} = \frac{\partial w_i c_i}{\partial z}- \frac{\partial}{\partial z} \left(D \frac{\partial c_i}{\partial z} \right),
\end{equation}
where the subscript $i$ can be $h$ or $l$ for heavy and light components respectively; $w_i$ is the segregation velocity (see $\S$\ref{SUBSECT: Segregation velocity}); and $D$ is the diffusion coefficient. Note that, because the bulk velocity is negligible in the $z$-direction, advection was neglected.

For what concerns diffusion, it is due to random particle collisions and it depends on local shear rate and on particle diameter \cite{Utter2004, Tripathi2013}. The scaling expression we used for $D$ reads: 
\begin{equation}
D=b \dot{\gamma} {D_p}^2,
\label{EQ: Diffusivity}
\end{equation}
where $b$ is a constant and it is assumed to be equal to 0.041 \cite{Utter2004}.
It is worthwhile underlining that the dependences of $D$ on the overburden pressure was proven to be negligible in the case of both size and density bi-dispersed flow \cite{Fry2019}.

To characterize the shear rate, the velocity profile of the particles as a function of the flow depth is required. In an annular shear cell, for a monocomponent material, the velocity profile decays exponentially with the bed height \cite{May2010, Artoni2018} so that it is well described by an exponential profile of the type:
\begin{equation}
v(z)=v_1+(v_0-v_1) \exp{\left(-\frac{z}{\delta}\right)},
\label{EQ: velocity profile}
\end{equation}
where $v$ is the particles horizontal velocity, $v_0$ and $v_1$ correspond respectively to the velocity of the bottom wall and the slip velocity at the top boundary; $\delta$ is the coefficient of the exponential decay which depends linearly on particle size. \citep{Artoni2018}. Here we assume that the velocity profile remains exponential also for density bidisperse mixtures.

Owing to the non-uniformity of the velocity gradient, the shear stress is obviously not constant throughout the domain. To obtain the shear rate, the velocity profile is differentiated as: 
\begin{equation}
\dot{\gamma}(z)=\frac{dv}{dz}=\frac{d}{dz} \left[ v_1+(v_0-v_1)\exp\left(-\frac{z}{\delta}\right)  \right].
\label{EQ: shear rate}
\end{equation}
Since the velocity profile is approximately time-independent, except for an initial transient \cite{May2010}, we do not consider the shear rate profile varying with time in segregation development.

At top and bottom walls, the no-flux boundary condition was imposed:
\begin{equation}
D \frac{\partial c_i}{\partial z}- w_i=0.
\label{EQ: boundary condition}
\end{equation}
The initial condition must correspond to the initial experimental configuration, so the heavy particle concentration in space, at time $t=0$, is represented with the following step function: 
\begin{equation}
    c_h(z,0) = 
    \begin{cases}
      0 & 0\leq z\leq \bar{z} \\
      1 & \bar{z}\leq z\leq H.
    \end{cases}
\label{EQ: initial conditions}
\end{equation}
where $\bar{z}$ is the location of the interface between light and heavy components.
For practical purposes, the concentration jump across the interface was smoothed with a second-order continuous smoothing function. This guarantees the continuity of the first and second order derivatives and hence, avoids discontinuous solution. 

\subsection{Segregation velocity}\label{SUBSECT: Segregation velocity}
Let's consider a heavy spherical particle of size $D_p$ and solid density $\rho_h$ in a medium composed of lighter particles. At steady state, the force of gravity is balanced by the buoyancy and the drag forces:
\begin{equation}
\bs{F_g}=\bs{F_b}+\bs{F_d} ,
\label{EQ: F_g}
\end{equation}
If in Eq. \ref{EQ: F_g}, the drag term is modelled as a viscous one, we obtain:
\begin{equation}
\frac{\pi}{6} D_p^3 \rho_h \bs{g} =\frac{\pi}{6} D_p^3 \rho_{m,s} \bs{g} + 6 \pi \eta \frac{D_p}{2} w_h ,
\label{EQ: force balance}
\end{equation}
where $\bs{g}$ is the gravitational acceleration that acts in the direction of fall and it is equal to $9.81$ m/s$^2$, and $\rho_{m,s}$ is the local density of the solid mixture. For a binary mixture of different density particles, $\rho_{m,s}$ is obtained weighting the solid density of each component by its local volumetric fraction as:
\begin{equation}
\rho_{m,s}=\rho_h c_h + \rho_l c_l .
\label{EQ: rho_m}
\end{equation}
After substituting Eq. \ref{EQ: rho_m} in Eq. \ref{EQ: force balance}, solving Eq. \ref{EQ: force balance} for the velocity gives: 
\begin{equation}
w_h=\frac{D_p^2 \bs{g}}{18 \eta} \left[ \left(1-c_h\right)\left(\rho_h - \rho_l \right) \right],
\end{equation}
which is analogous to the terminal velocity of a sphere that falls under gravity in a viscous medium as predicted by the Stokes' law \cite{Bird2002}. 
Thus, the segregation velocities of heavy and light particles under shear in a dense bi-dispersed granular flow are respectively equal to:
\begin{equation}
w_h=\frac{D_p^2 \bs{g} }{18 \eta} \left[\rho_h \left(1-c_h\right)\left(1 - \frac{1}{\delta_R} \right) \right],
\label{EQ: w_h}
\end{equation}
\begin{equation}
w_l=\frac{D_p^2 \bs{g} }{18 \eta} \left[\rho_l \left(1-c_l\right)\left(1 -  \delta_R\right) \right].
\label{EQ: w_l}
\end{equation}
It should be noted that the higher the density difference, the higher the driving force leading to segregation. Moreover, segregation is faster when a particle is surrounded by a larger amount of grains of the other component. The dependence of the segregation velocity on shear rate and pressure gradient is discussed below (see $\S$ \ref{SUBSECT: constitutive laws}).

\subsection{Constitutive relations}\label{SUBSECT: constitutive laws}
To provide a closure for the transport equation, a constitutive relation for the effective viscosity, $\eta$, has to be introduced in the model.

The effective viscosity, $\eta$, is commonly modelled with the law proposed by Jop et al. \cite{Jop2006} that reads: 
\begin{equation}
\eta(|\dot{\gamma}|,P)=\frac{\mu(I) P}{|\dot{\gamma}|}.
\label{EQ: viscosity law}
\end{equation}
This effective viscosity depends on both shear rate and local pressure \cite{Jop2006}.
The latter evolves in time and varies with depth according to:
\begin{equation}
P(z,t)=P_{load}+g\int_{h}^{H} \rho_b(z,t) dz .
\end{equation}
The definition of granular viscosity (Eq. \ref{EQ: viscosity law}) requires also a closure relation for the friction coefficient $\mu(I)$. The friction coefficient increases with the inertial number, $I$, from a minimum value, $\mu_{min}$, at zero shear rate, and saturated to $\mu_{2}$ at higher values of $I$. This because, granular matter shares similarities with viscous-plastic fluid, such as Bingham fluids \cite{Jop2006}.
As friction law, we used the one proposed by Jop et al. \cite{Jop2006} that reads:
\begin{equation}\label{EQ: mu(I)}
	\mu(I)=\mu_{min}+\frac{\mu_2-\mu_{min}}{I_0/I+1}
	\end{equation}
where $\mu_{min}=\tan(20.90^o)\approx0.38$, $\mu_{2}=\tan(32.76^o)\approx0.64$ and $I_0=0.279$ for glass spheres.
The inertial number $I$ is defined as:
\begin{equation}
I \equiv |\dot{\gamma}| D_p \sqrt{ \frac{\rho_{m,s} \phi_s}{P} }.
\end{equation}
{The solids volume fraction, $\phi_s$, was assumed constant and approximately equal to 60\%. The value 60\% is obtained from experimental evidences at the beginning of the tests. Instead, the assumption of being constant is reasonable since the bed is confined by an overload and, unlike size segregation, the local porosity does not change accordingly to the local concentration of each specie.}

\textcolor{black}{
We want to highlight that the scaling law in Eq. \ref{EQ: mu(I)} has been empirically determined by Jop et al. \cite{Jop2006} with experiments carried out using glass beads having $0.53$ mm diameters. However, in our experiments we employed particles with larger size, made of different materials and thus, with likely different frictional properties. Furthermore, our fully confined flow configuration surely influences the macroscopic rheology. 
To account for these effects in the rheological model, the friction coefficient was simply corrected with a constant factor, $k$.}
The corrected friction coefficient becomes:
\begin{equation}
\mu^*(I) =  k \cdot \mu(I).
\end{equation}
By replacing $\mu(I)$ with $\mu(I)^*$ in Eq. \ref{EQ: viscosity law}, we obtained the expression for the viscosity $\eta$:
\begin{equation}
\eta(I) = \frac{\mu^*(I) P}{|\dot{\gamma}|}
\end{equation}
The optimal value of $k$ is the one that leads to the best match between experimental and simulated profiles.

Notice that the use of the rheological model regarded only the definition of the segregation law, which contains the solid viscosity, but did not enter in the calculation of the flow field, which was assumed to be known. 

\section{Experimental results}\label{SECT: Result_exp}
\begin{figure} \centering
	\includegraphics[width=\mysize\textwidth]{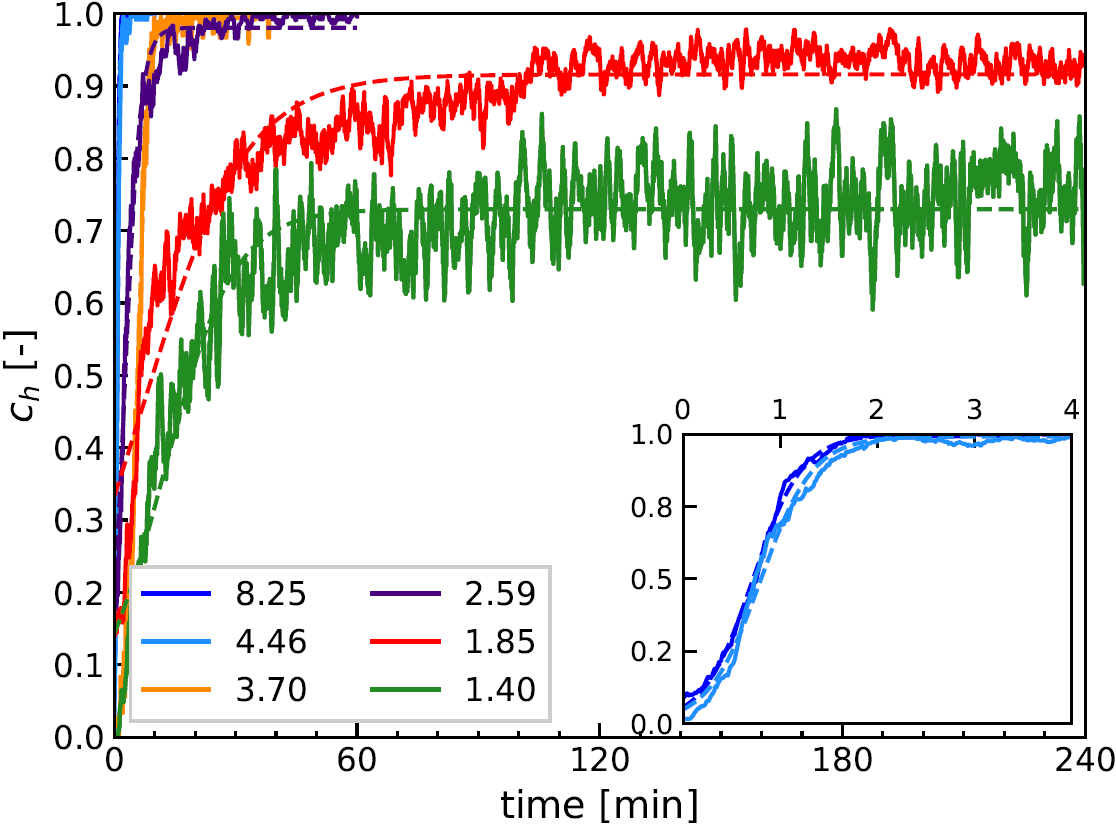}
	\caption{Evolution of the concentration profiles of the heavy grains at the bottom. The profiles are shown for different density ratio $\delta_R$. In all cases the experiments were started by filling the bottom-half of the cell with the lighter material, and the top-half of the cell with the denser material (50\% initial overall concentration). In the inset, the 8.25 and 4.46 density ratio curves are represented on a larger scale. Note that all solid lines represent the experimental results, whereas the dashed lines are fitted with Eq. \ref{EQ: fitting_exp_bottom}.}
	\label{FIG: exp_bottom}
	\end{figure}

As a first macroscopic assessment of the evolution of segregation, we analysed the concentration profiles of the heavy component in the bottom part of the cell. The assessment was done for six different density ratios and 50\% volumetric fraction. The resulting profiles are displayed in Fig. \ref{FIG: exp_bottom}. 
When the granular material flows, all particles start diffusing and segregating: heavy grains percolate downward, whereas light particles rise through the bed due to the buoyant force. The heavy particle concentration grows from 0, passes through 50\% concentration, and then reaches the steady state. Such a system has moved from the initially segregated configuration to a well-mixed state, until re-segregating again. 
In the case of higher density ratios, the driving force leading to segregation is higher and dominates over diffusion. For smaller density ratios, diffusion gains always more importance over segregation.
For this reason, the lower the $\delta_R$, the lower the final $c_h$ reached (e.g. $c_{h,b}$ settles at around 73\% for $\delta_R=1.40$). 
In order to extract some parameters describing the process dynamics, the experimental profiles were fitted by solving the Least Squares Minimization problem with the following hyperbolic tangent function:
\begin{equation}
f(t)=\frac{A}{2} \cdot \left[ 1- \tanh \left( - \frac{t-t_0}{\tau -t_0} \right) \right]. 
\label{EQ: fitting_exp_bottom}
\end{equation}
where $t_0$ and $\tau$ are respectively the time required to reach the 50\% and the 88\% of the final heavy particle concentration, named $A$. The fitting curves are represented as dashed lines in Fig. \ref{FIG: exp_bottom} together with the raw data.

\begin{figure} \centering
	\includegraphics[width=\mysize\textwidth]{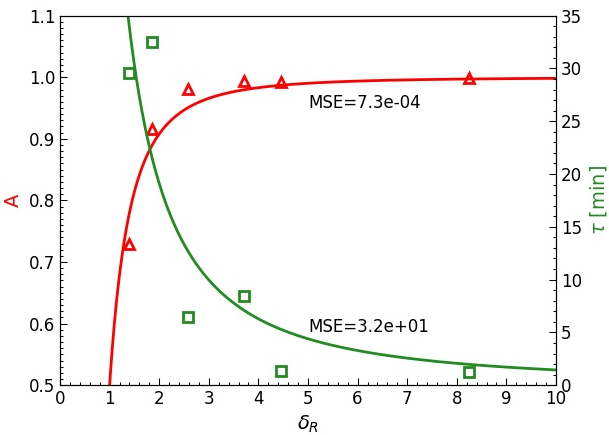}
	\caption{Plot of the fitting parameters $A$ and $\tau$ as a function of the density ratio for $c_h=50\%$. The first parameter refers to the left yaxis, whereas the latter refers to the right yaxis. The Mean Squared Errors (MSE) are also reported.}
	\label{FIG: parameters_exp_bottom}
	\end{figure}

In Fig. \ref{FIG: parameters_exp_bottom}, the parameters $A$ and $\tau$ and their fitting curves are reported as a function of density ratio. $A$ can be estimated with good accuracy (Mean Squared Error equal to $7.3\cdot 10^{-4}$) with the following function: 
\begin{equation}
A=1-\frac{1}{2\cdot \delta_R^{\beta}} \quad \delta_R>1,
\label{EQ: fitting parameter A}
\end{equation}
where $\beta=2.27$ \textcolor{black}{is an exponent characterizing the sensitivity of the final segregation state on the density ratio.} For what concerns $\tau$, it follows an exponential decay but, as the $MSE$ reveals, its fitting is not so accurate.
In the limit of $\delta_R=1$ (i.e. when the two species have the same density), the system is always perfectly mixed ($c_h=50\%$) and segregation never occurs (diffusional mixing prevents separation). On the other hand, when $\delta_R$ becomes very large, the time required to reach the re-segregated state tends to 0, and segregation happens almost instantaneously. 

\begin{figure}[htbp]\centering
\includegraphics[width=\mysize\textwidth]{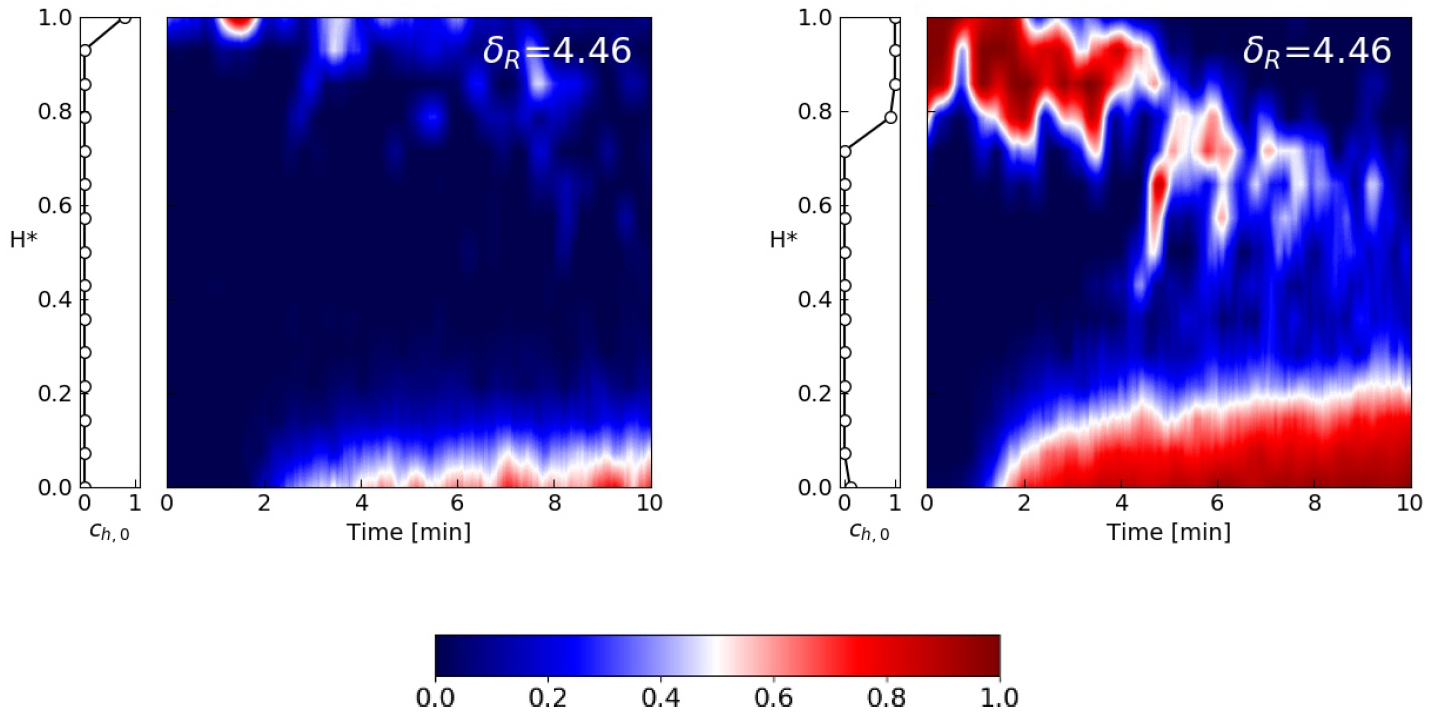}
	\caption{Experimental measurements of the heavy particle concentration distributions in time \textcolor{black}{and as a function of the dimensionless cell height, $H^*$} for $c_{h,0}=10\%$ (on the left) and $c_{h,0}=25\%$ (on the right). In both cases, $\delta_R=4.46$. The colour bar refers to $c_h$. \textcolor{black}{At the left of each contour plot, the initial concentration profile is also displayed.}}
	\label{FIG: exp 0.25}
	\end{figure}

\begin{figure}[h!]\centering
\includegraphics[width=\mysize\textwidth]{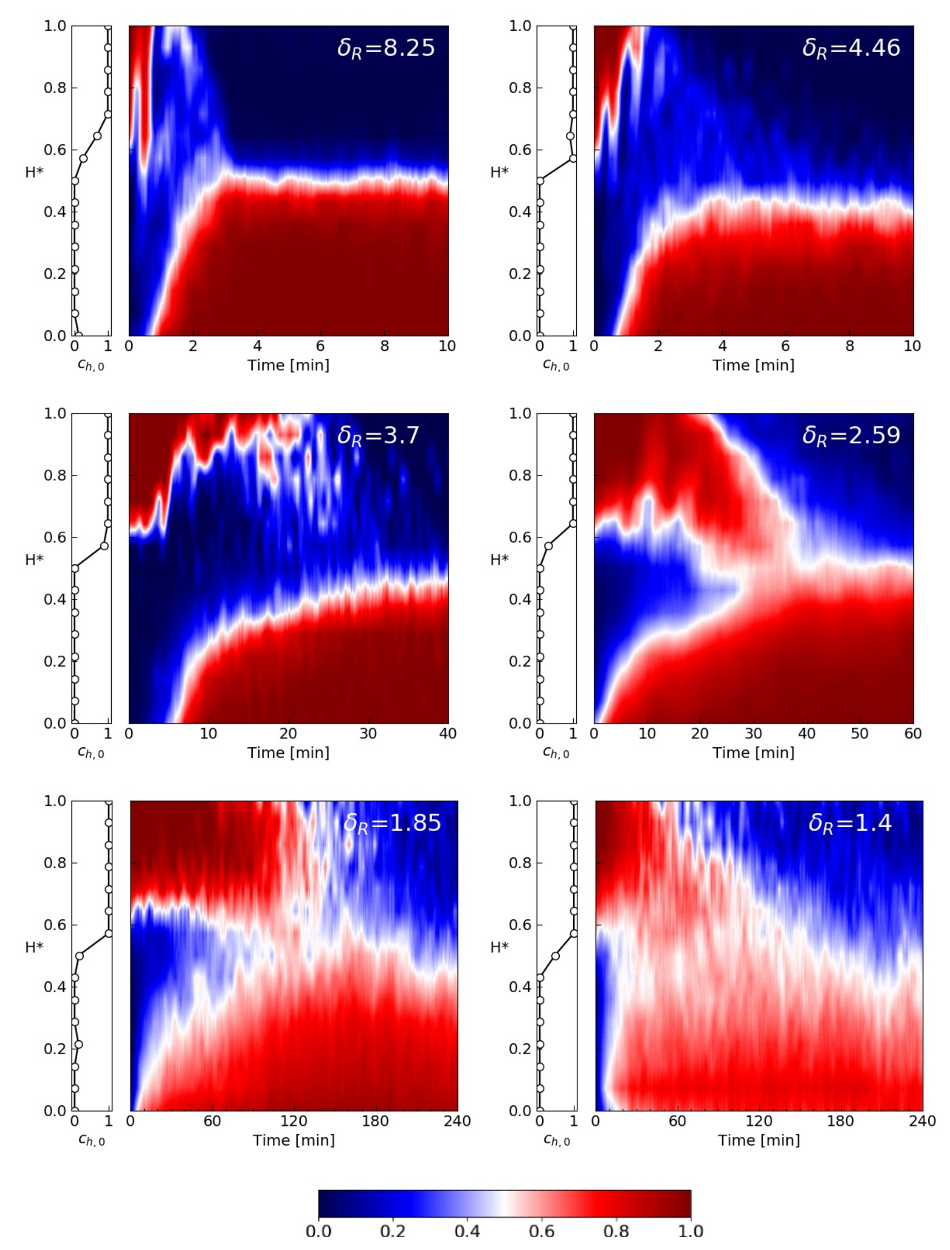}
	\caption{Experimental measurements of the heavy particle concentration distributions in time for $c_{h,0}=50\%$ at decreasing density ratio. \textcolor{black}{At the left of each contour plot, the initial concentration profile is also displayed.}}
	\label{FIG: exp 0.50}
	\end{figure}

\begin{figure}[h!]\centering
	\includegraphics[width=\mysize\textwidth]{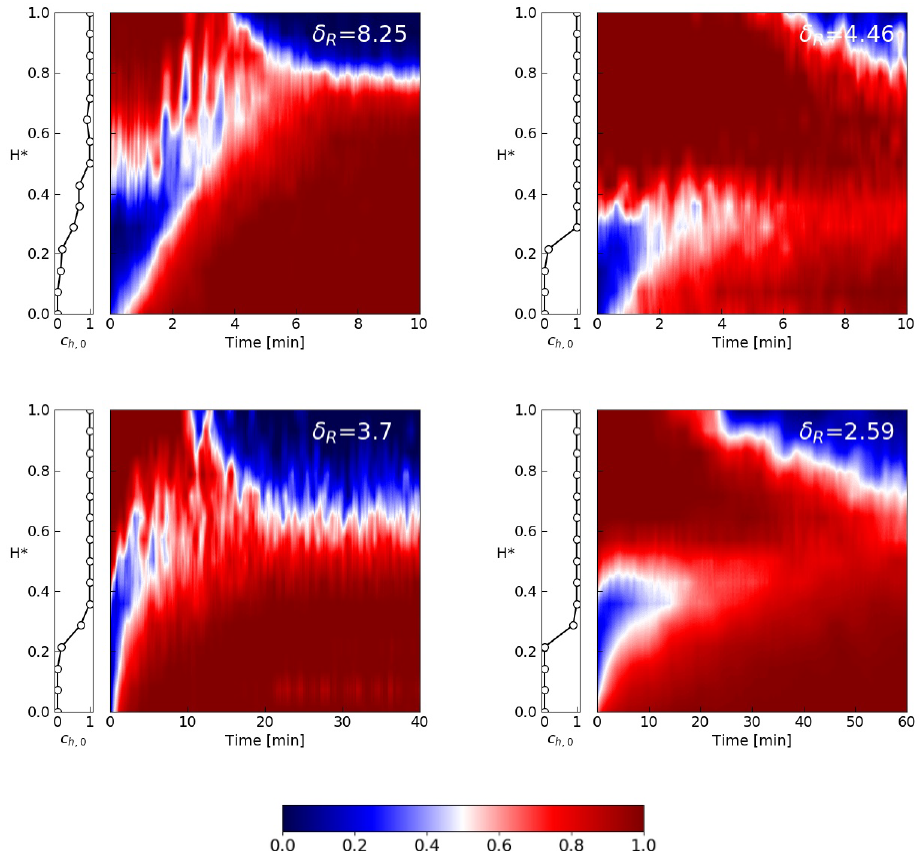}
	\caption{Experimental measurements of the heavy particle concentration distributions in time for $c_{h,0}=75\%$ at decreasing density ratio. \textcolor{black}{At the left of each contour plot, the initial concentration profile is also displayed.}}
	\label{FIG: exp 0.75}
	\end{figure}

\begin{figure}[h!]\centering
\includegraphics[width=\mysize\textwidth]{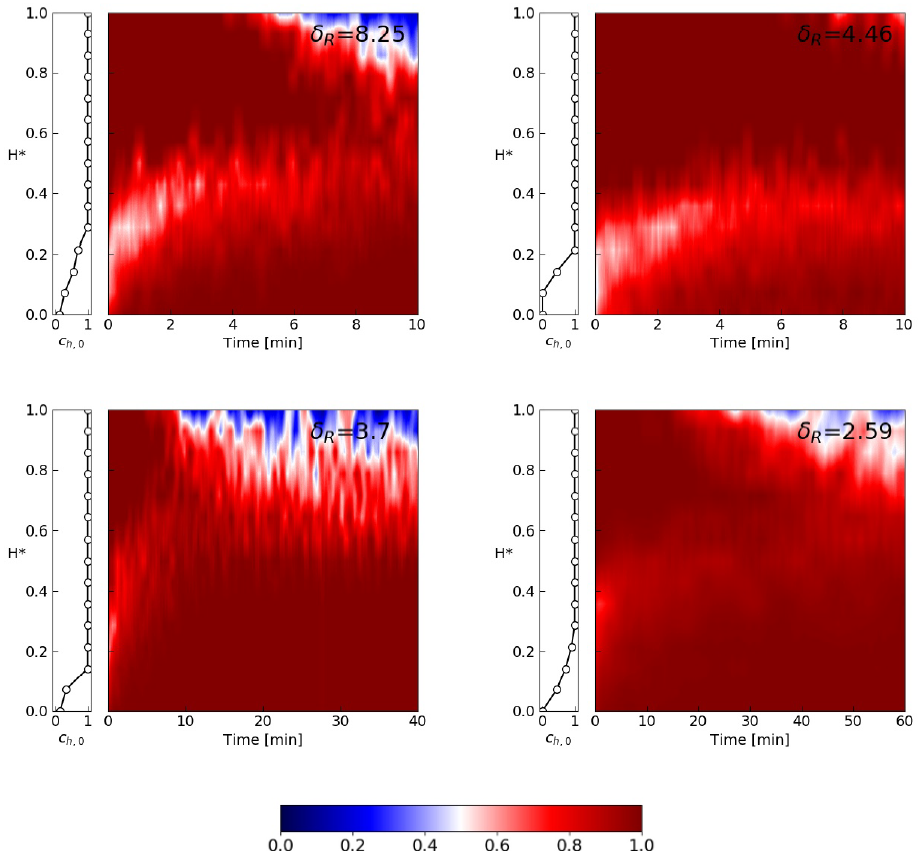}
	\caption{Experimental measurements of the heavy particle concentration distributions in time for $c_{h,0}=90\%$ at decreasing density ratio. \textcolor{black}{At the left of each contour plot, the initial concentration profile is also displayed.}}
	\label{FIG: exp 0.90}
	\end{figure}

To see in more detail how heavy particle concentration evolves in the entire flow depth, we have reconstructed the temporal evolution of the heavy particle concentration through the depth of the cell. The resulting contour maps are reported in Fig. \ref{FIG: exp 0.25} to Fig. \ref{FIG: exp 0.90}, respectively for increasing volume fractions. The contour maps are characterised by 100 contour regions and, to reduce the noise, concentrations have been smoothed over time by applying the moving average.
Independently of $\delta_R$ and $c_{h,0}$, the time required by the heavy particles to reach the bottom is always smaller than the time required by the smaller particles to reach the top. Furthermore, the final fully segregated state is reached faster in the case of higher $\delta_R$ and for $c_{h,0}=50\%$. It is interesting to see that, the time required to reach the steady state is higher for $\delta_R=1.85$ than for $\delta_R=1.40$, despite the higher density ratio, being  $c_{h,0}=50\%$ equal. This is because the total mass of grains was $M_g=1.62$ kg and $M_g=2.52$ kg for $\delta_R=1.85$ and $\delta_R=1.40$ respectively. As we will see, the lower the mass under the same load, the thinner the shear band.

At the interface particles partially mix due to diffusion. 
This indicates that diffusive remixing due to the random collisions between the grains as they are sheared competes against density-driven segregation. This competition is relevant especially for small $\delta_R$, whereas for higher $\delta_R$ the interface is sharper.
\section{Numerical results}\label{SECT: Result_num}
Rather than deforming uniformly, granular materials under shear stresses develop shear bands, zones of intense shear close to essentially rigid regions \cite{Mueth2000}. The formation of shear bands is often explained by the existence of a yield shear stress \cite{Artoni2018}. Very little is known about shear band and about how the microstructure of individual grains affects movement in densely packed material \cite{Mueth2000}. Mueth et al. \cite{Mueth2000} found that at high packing density and slow shear rate, the key characteristics of the granular microstructure determine the shape of the velocity profile. Artoni et. al \cite{Artoni2018} demonstrated that the exponential decay appearing in the velocity profile ($\delta$) is strictly related to the ratio between the mass loaded on the system and the mass of the grains: $\tilde{M}\equiv M_w/M_g$.

For this reason, we gave verification of the segregation model proposed in $\S$ \ref{SECT: Num. model} for three cases having a similar $\tilde{M}$, as reported in Tab. \ref{TAB: testes case numerically}.
\begin{table}[h]
	\centering \footnotesize
	\begin{tabular}{lcccc}
	\textbf{}     & $c_{h,0}$ & \textbf{Mg} & \textbf{Mw/Mg} & \textbf{Mw+Mg} \\ \hline
	C-PP    & 50\% & 2.67        & 0.41           & 3.77           \\
	G-PP    & 50\% & 2.04        & 0.54           & 3.14           \\
	G-POM   & 50\% & 2.52        & 0.44           & 3.62           \\ \hline
	\end{tabular}
	\caption{The segregation model has been tested for these three cases that are characterized by a similar $\tilde{M}=M_w/M_g$ [-]. \textcolor{black}{$M_g$ and $M_w$ are expressed in kg.}}
	\label{TAB: testes case numerically}
	\end{table}

The model was implemented in a finite element commercial code (COMSOL$\text{\textregistered}$).
For each test, simulations were run for different values of the parameter $k$ and different values of the exponential decay $\delta$. The test were planned as per full factorial design. 
Following \cite{Artoni2018}, the value of $\delta/H$ should be close to 0.20 and hence, we chose to test  $\delta/H=0.18$, $0.20$ and $0.22$. Furthermore, preliminary test showed that a value $k=0.60$ is reasonable to capture experimental results. Thus, we performed simulations with $k=0.5$, $0.6$ and $0.7$.  
We used as key performance indicator, $KPI$, the root-mean-square deviation between experimental and simulated contour maps. The evaluation of the $RMSD$ between contour plots (i.e. matrices) was calculated as:
\begin{equation}
RMSD=\sqrt{\frac{\sum^{m}_{i=1} \sum^{n}_{j=1} (c_{ij, exp}-c_{ij, num})^2}{m\cdot n}} ,
\end{equation}
where $m$ and $n$ are the matrices dimensions. The optimal combination of values $k$ and $\delta$ is the one that minimize the $RMSD$.

Tab. \ref{TAB: Full Factorial Design} lists, for each combination of the factors, the resulting $RMSD$ in the case of 50:50 mixtures of 1) Ceramic and Polypropylene, 2) Glass and Polypropylene and 3) Glass and filled-POM. 
It is clear that, the combination of values that minimize the RMSD is $k=0.7$ and $\delta/H=0.18$.

\begin{table}[h]
	\centering \footnotesize
	\begin{tabular}{ccc|ccc}
	 & \multicolumn{2}{l}{\textbf{Factors}} & \multicolumn{3}{c}{\textbf{RMSD}}                                                                                    	\\ \hline
	\textbf{ID}  & \textbf{k}  & \textbf{$\delta/H$}  & \multicolumn{1}{l}{\textbf{C-PP}} & \multicolumn{1}		{l}{\textbf{G-PP}} & \multicolumn{1}{l}{\textbf{G-POM}} \\ \hline
	\textbf{1}   & 0.50        & 0.18 & 0.206 & 0.217 & 0.149                                   \\
	\textbf{2}   & 0.50        & 0.20 & 0.254 & 0.252 & 0.162                                   \\
	\textbf{3}   & 0.50        & 0.22 & 0.283 & 0.277 & 0.170                                   \\
	\textbf{4}   & 0.60        & 0.18 & 0.182 & 0.194 & 0.130                                   \\
	\textbf{5}   & 0.60        & 0.20 & 0.231 & 0.227 & 0.142                                   \\
	\textbf{6}   & 0.60        & 0.22 & 0.262 & 0.252 & 0.150                                   \\
	\rowcolor[HTML]{C0C0C0} 
	\textbf{7}   & 0.70        & 0.18 & 0.169 & 0.182 & 0.119                                   \\
	\textbf{8}   & 0.70        & 0.20 & 0.213 & 0.209 & 0.130                                   \\
	\textbf{9}   & 0.70        & 0.22 & 0.246 & 0.233 & 0.138 \\ \hline
	\end{tabular}
	\caption{The results obtained from the full factorial design of experiment for: 1) C-PP, 2) G-PP and 3) G-POM.}
	\label{TAB: Full Factorial Design}
	\end{table}

\begin{figure}[h!]\centering
\includegraphics[width=\mysize\textwidth]{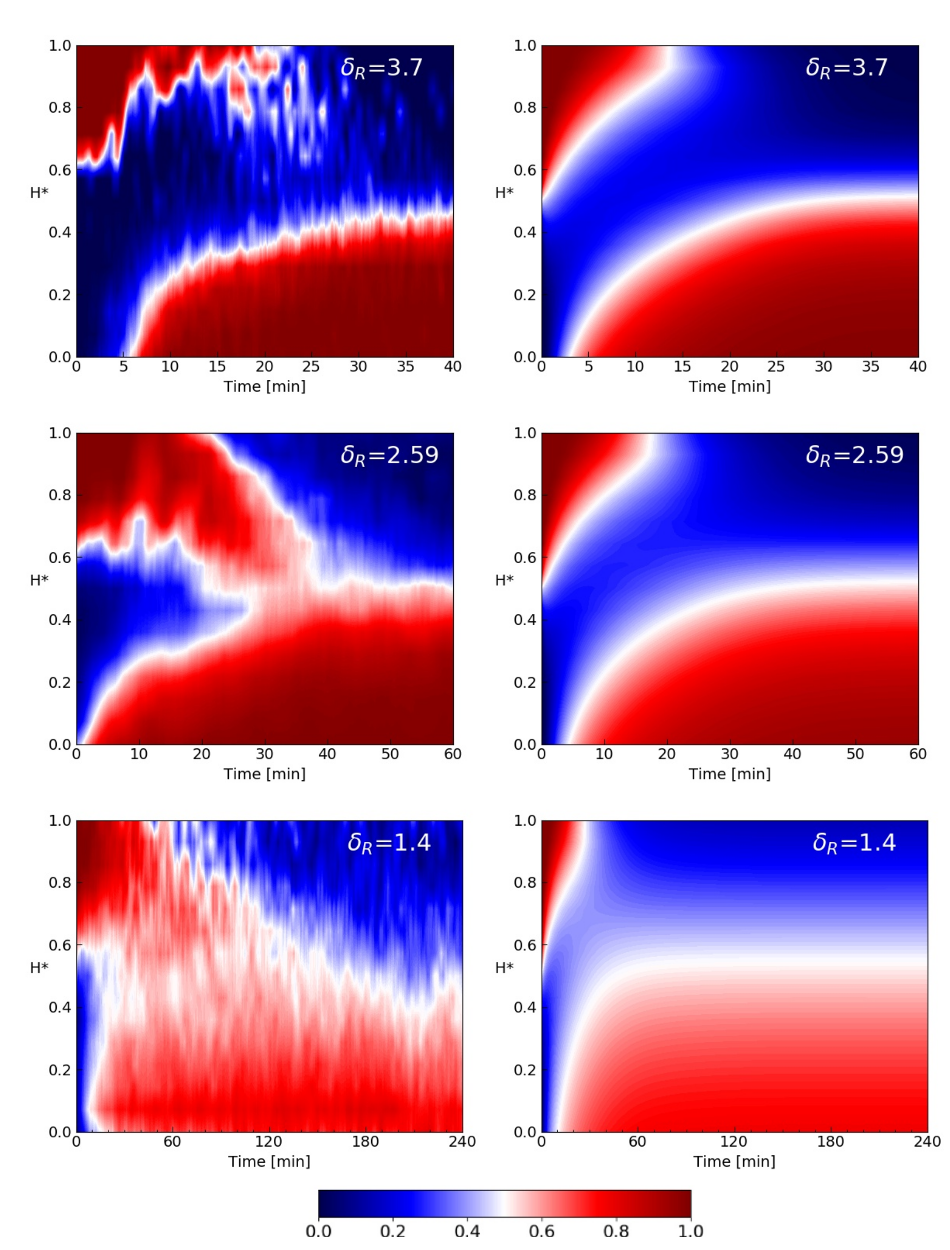}
	\caption{Comparison between experimental finding (first column) and numerical results (second column) for $c_{h,0}=50\%$. The numerical contour maps have been obtained with the optimal combination of $k$ and $\delta$. The RMSD are: 0.169, 0.182 and 0.119 for $\delta_R$ equal to $3.70-2.59-1.40$ respectively.}
	\label{FIG: exp num comparison 0.50}
	\end{figure}

\begin{figure}[htbp]\centering
\includegraphics[width=\mysize\textwidth]{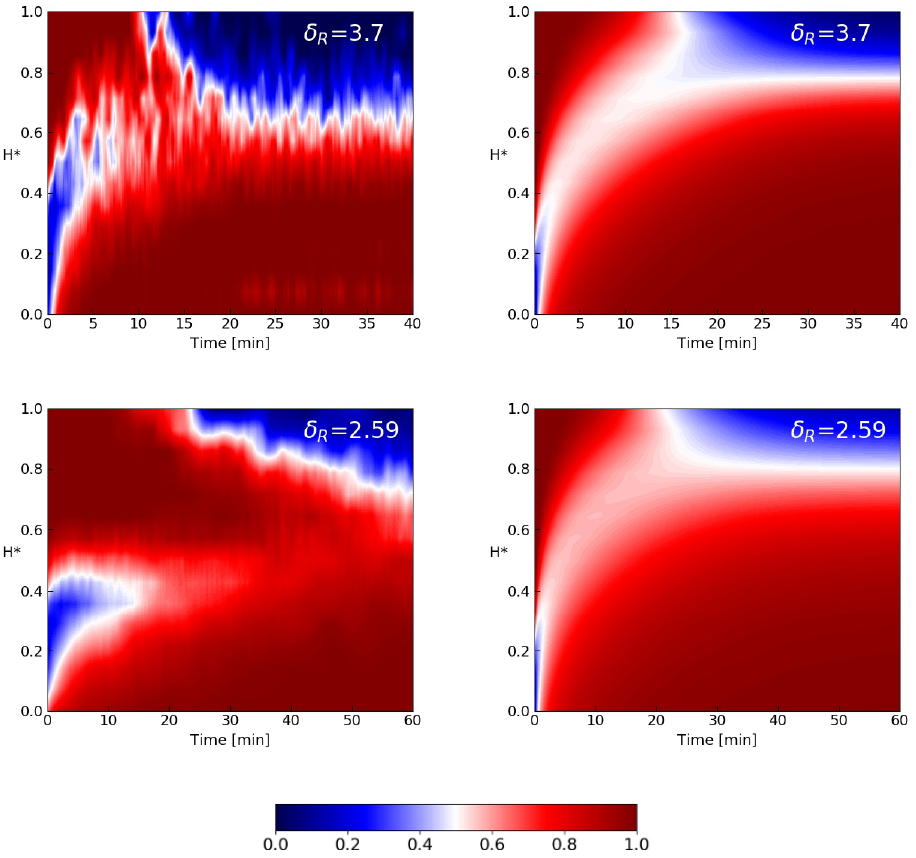}
	\caption{Comparison between experimental finding (first column) and numerical results (second column) for $c_{h,0}=75\%$. The numerical contour maps have been obtained with the optimal combination of $k$ and $\delta$. The RMSD are: 0.166 and 0.176 for $\delta_R$ equal to $3.70$ and $2.59$ respectively.}
	\label{FIG: exp num comparison 0.75}
	\end{figure}

\begin{figure}[htbp]\centering
	\includegraphics[width=\mysize\textwidth]{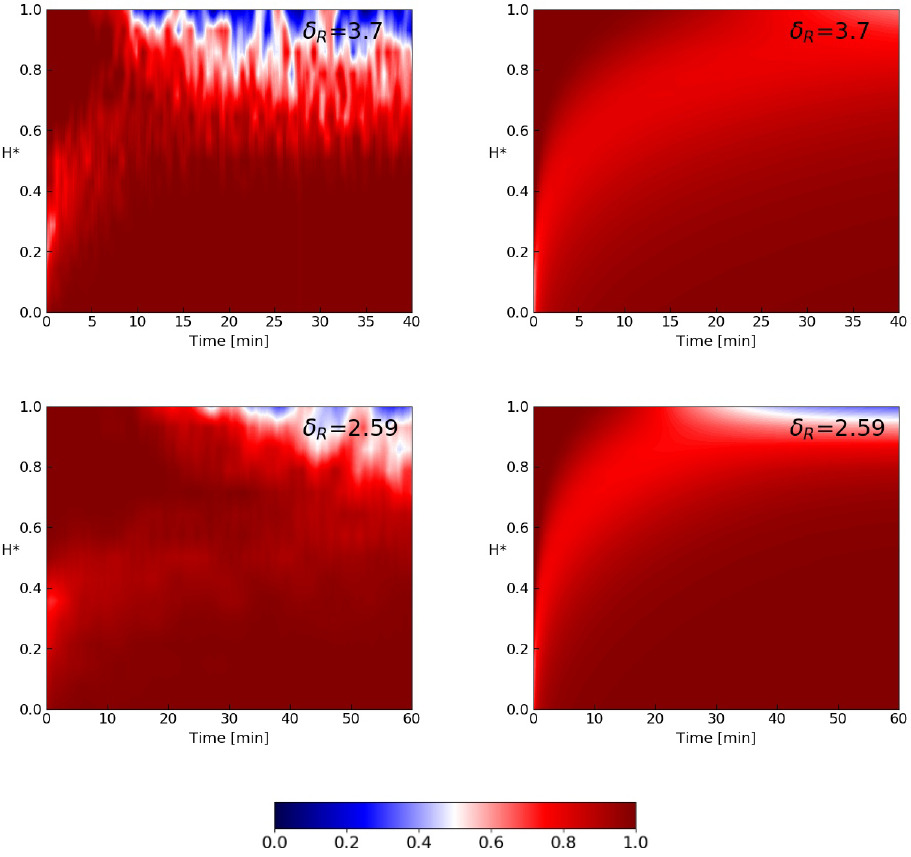}
	\caption{Comparison between experimental finding (first column) and numerical results (second column) for $c_{h,0}=90\%$. The numerical contour maps have been obtained with the optimal combination of $k$ and $\delta$. The RMSD are: 0.175 and 0.091 for $\delta_R$ equal to $3.70$ and $1.40$ respectively.}
	\label{FIG: exp num comparison 0.90}
	\end{figure}

In Figs. \ref{FIG: exp num comparison 0.50}, \ref{FIG: exp num comparison 0.75} and \ref{FIG: exp num comparison 0.90} the comparison between experimental findings and numerical results, computed with the optimal combination of $k$ and $\delta$, are reported for 50\%, 75\% and 90\% volumetric fraction of the heavy component, respectively. 
As expected, the experimental results are much noiser than the numerical ones due to the discrete nature of our system and the continuous approach used in the simulations. However, since the $RMSD$ between experimental and numerical outcomes is always smaller than 18\%, we can conclude that they agree both qualitatively and quantitatively. The proposed model is therefore able to captures the main features of density-driven segregation in the case of densely packed flow, also for non homogeneous shear rates. 
However, it should be noted that $\delta$ is a function of $\tilde{M}$ therefore, $\delta=0.18$ works for the cases reported in Tab. \ref{TAB: testes case numerically}. We will prove that the dynamics of the granular flow, and hence the features of segregation, is strictly connected to $\delta$ in section $\S$\ref{SECT: discussion}. 

During the mixing stage, we experimentally saw (see Fig. \ref{FIG: exp num comparison 0.50}) more heavy particles (red) on the external side-wall than numerically. This effect becomes also more remarkable for higher $\tilde{M}$, suggesting that the applied load influences not only the shear features, but also the transverse wall friction coefficient \cite{Artoni2018}. For lower $\tilde{M}$, the shear band is larger, the transverse wall friction coefficient is definitely lower than its streamwise counterpart and hence, the horizontal flow of grains dominates over radial motion. On the other hand, for higher values of $\tilde{M}$, the shear band is thinner, the radial forces becomes always more important and particles starts moving significantly also in the radial coordinate. This may lead to a three-dimensional recirculation pattern: while rising, small particles moved away from the outer wall for reappearing once reached the top. 
Since our model is one-dimensional, it cannot capture this feature and so, we always see the small particles at the outer surface. 
It should be noted that, the presence of this three dimensional recirculation pattern in geometries similar to our, has always been neglected or minimized in the existing literature, also in the case of size-driven segregation. Let's see for instance at Fig. 2 of L. May et al. (2010) \cite{May2010}: while rising, not all the large particles appears at the window in the outer wall, indicating that something is hidden in the third dimension. A detailed study on the relative influence of the effective streamwise and transverse wall friction coefficients on granular flow is required and it will be the subject of a future contribution.


\section{Shear localization and velocity profile}\label{SECT: discussion}
Artoni et al. \cite{Artoni2018} show that the velocity profile is not universal but depends on the flow parameters. Furthermore, the exponential decay, $\delta$, is affected by the joint effect of normalized applied load, $\tilde{M}$, and flow depth. In the limit of small $\tilde{M}$, the pressure at the bottom is low with respect to that induced by the grains, and the shear localizes in a wider band. Also a decrease in the flow depth acts in the same way.
In our cases, the flow was always 15 particle diameters depth, therefore it can be assumed that the normalized applied weight, $\tilde{M}$, is the only factor influencing $\delta$. 

To prove that the dynamics of granular flow is strictly related to the height of the shear band, which in turn is linked to the normalized applied weight, we performed velocity profile measurements.
The cell was completely filled with one material at a time, except for the steel because too heavy to be supported by the motor. 
The operating conditions were the same as for the segregation experiments: the bottom wall was rotating at 23.44 rpm and 1.093 kg was loaded on the top. 
Differently from before, the process was recorded with a high-speed camera (Phantom Miro 320S). For each material, three movies were recorded at 24, 100 and 1000 frames per second, respectively. Working with different frame-rates is indeed required to obtain meaningful profiles for all the flow depth.
The videos were processed with the free software ImageJ. The displacement of particles between consecutive frames was manually tracked and the velocity profile reconstructed by connecting the segments obtained at different frame rates. 

We also carried out discrete element method (DEM) simulations to obtain the velocity profiles of glass and steel, whose simulation details are reported in appendix. Since the velocity profile of glass is in agreement with the experimental one, our simulation is quantitatively valid and the simulated profile of steel, which was impossible to determine experimentally, can be considered trustworthy.

\begin{figure}
	\centering
	\includegraphics[width=\mysize\textwidth]{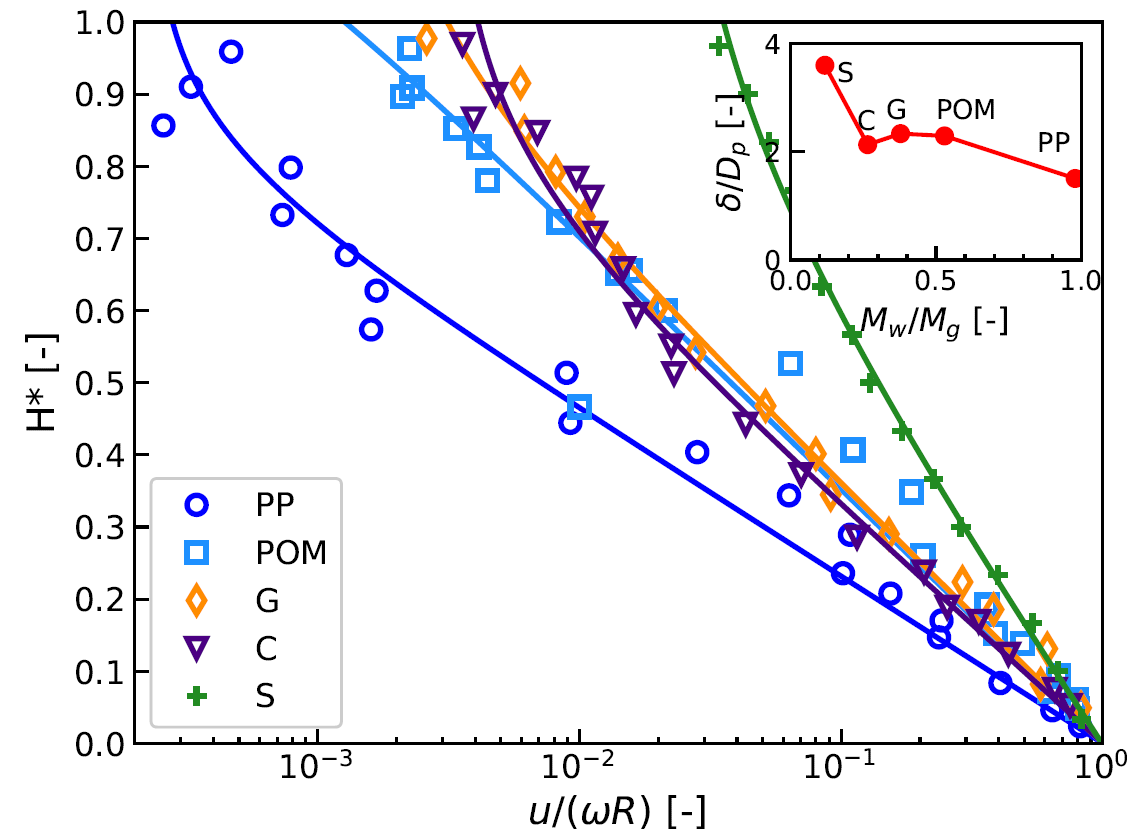}
	\caption{Velocity profile for the pure components. The profile for PP, POM, G and C were obtained from experiments, whereas the one of steel by DEM simulation.}
	\label{FIG: velocity profile}
	\end{figure}

Fig. \ref{FIG: velocity profile} shows the velocity profile thus obtained. The decay is steeper for Polypropylene, whereas the degree of decay is lower for steel. The difference in the decay of the velocity profile is in agreement with what is reported in \citep{Artoni2018} and is due to the different masses of the grains. In the inset of Fig. \ref{FIG: velocity profile}, the dimensionless coefficient of exponential decay is reported as a function of $M_w/M_g$. As expected, this shows that steel is characterized by a wider shear band than lighter material under the same load, at least for 15 particle diameters flow depth.

\begin{figure}
	\centering
	\includegraphics[width=\mysize\textwidth]{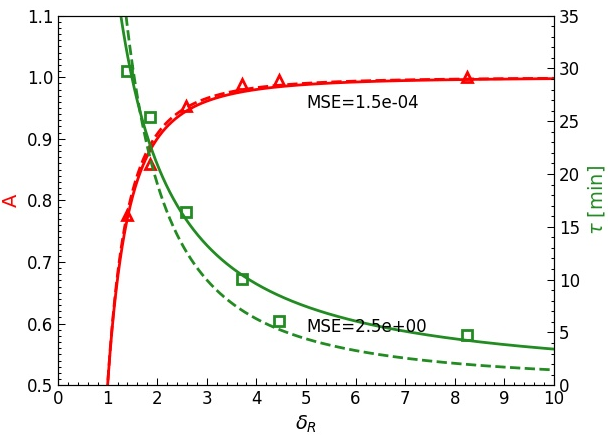}
	\caption{Plot of the fitting parameters A and $\tau$ that we obtained numerically with $k=0.70$ and $\delta=0.18$ as a function of the density ratio for $c_h=50\%$. The first parameter refers to the left yaxis, whereas the latter refers to the right yaxis. The Mean Squared Errors (MSE) are also reported. For comparison, we have drawn also the fitting of the experimental parameters (dashed lines).}
	\label{FIG: A tau numerical}
	\end{figure}

Fig. \ref{FIG: A tau numerical} reports the values of the parameters $A$ and $\tau$ that we achieved numerically for a range of density ratios and considering $k=0.70$ and $\delta=0.18$. The fitting of A is done with  Eq. \ref{EQ: fitting parameter A} and $\beta=2.32$. Since the experimental $\beta$ was $2.27$, there is good agreement between experiments and simulation. 
Nevertherless, $\tau$ is overestimated for $\delta_R=8.25$ and $\delta_R=4.46$, and underestimated for $\delta_R=1.85$. 
The cases with $\delta_R=8.25$ and $\delta_R=4.46$ are characterized by a greater $\tilde{M}$ and, the exponential decay must be higher than 0.18. In contrast $\delta_R=1.85$ is the case with the smallest $\tilde{M}$, the shear band is the thinnest, and $\delta$ must smaller than 0.18 in order to have reliable results.
It is now evident that, consistently with our main assumption, $\delta$ is sensitive to $\tilde{M}$.

\section{Conclusions}\label{SECT: conclusions}
The objective of the present investigation was to observe and model density-driven segregation in the case of dense granular flow where long-lasting contacts between neighbourhood particles are dominant. 

Firstly, we have performed quantitative experiments. The segregating system was characterized by a shear-rate gradient that lead to segregation of particles according to their density. The assessment was done for binary mixtures initially segregated, differing in density ratio and volumetric fraction. To the authors's knowledge, this is the first time that density driven segregation is experimentally studied in an annular shear cell where the flow is uninterrupted and where there is no need to feed material. 
The experimental results show that it is not always true that the higher the density ratio, the faster the segregation process: the segregation process is indeed strictly related to the flow features of the system such as the degree of the exponential decay of the velocity profile. We also show that, the higher the density ratio, the sharper the interface reached at the final steady state.  

A mathematical model has successively been developed. The proposed segregation velocity is analogous to the settling velocity predicted by the Stokes' law. It should be noted that a similar model has been found in a previous work of Tripathi and Khakhar \cite{Tripathi2013}. However, unlike them, we do not described granular segregation in terms of effective temperature and we do not combined the segregation model with a rheological model. In our analysis, which aimed to develop an expression for density-driven segregation velocity, the flow field was modelled using an exponential velocity profile. The optimal values of the two model unknowns where obtained by a full factorial design of experiment aimed at minimizing the RMSD between experimental finding and numerical outcomes. Overall, the results of the mathematical model match the results determined from experiments for different density ratios.

We found that, understanding the shear band localization is crucial to describe and predict granular flow onset and rheology \cite{Artoni2018}. We also shows that, in the limit of high normalized loaded mass, a three-dimensional flow pattern takes place within the granular bed, influencing what we see at the sidewall. In the view of the above, the segregation model gives quantitatively and qualitatively good results.

The main advantage of our continuum model is that it permits analytical and fast numerical solutions for a range of density ratios and different volumetric fractions. The model can therefore be applied broadly for investigating and designing binary powder systems. The only drawback is that it requires the knowledge of the velocity profile since the shear localization features influence the segregation time. 
This problem could be solved by coupling the density-segregation velocity to a non-local rheology.

Further studies are required in order to better understand the evolution in time of the velocity profile, the shear localization patterns and the effect of the wall friction coefficient on different mixture of grains since, all of these effects can affect segregation.

\newpage
\section*{Appendix: Simulation method} 
We implemented our 3D soft-sphere simulation with LIGGGHTS\textsuperscript{\textregistered}-PUBLIC  \cite{liggghts, liggghts_doc}, an open source DEM particle simulation software. 
In DEM simulations, the trajectory of each particle is calculated considering all the forces acting on it and integrating Newton’s second law of motion and the kinematic equations for position and orientation \cite{DiRenzo2004}. The translational and rotational motions are given by the following equations \cite{thornton2015}:
\begin{equation}
	\frac{d\bm{v}_i}{dt}= \frac{\sum \bm{F}_{ij}}{m_i}+\bm{g}_i
	\label{EQ: translational acc.}
	\end{equation}
\begin{equation}
	\frac{d\bm{\omega}_i}{dt}= \frac{\sum \bm{F}^T_{ij} R_i+ \bm{\tau}_{ij}}{I}
	\label{EQ: rotational acc.}
	\end{equation}
where $\bm{v}_i$ and $\bm{\omega}_i$ are the linear velocity and the angular velocity, $\bm{I}$ is the moment of inertia and $\bm{g}_i$ is the acceleration due to gravity of particle $i$. The force $\bm{F}_{ij}$ is the contact force between the grain $i$ and the grain $j$, whereas $\bm{F}^T_{ij}$ is its tangential component. By Netwon's third law, the force experienced by particle $j$ is: $\bm{F}_{ji}=-\bm{F}_{ij}$ \cite{Chialvo2012}.
If one wants to account for the effect of rolling friction during contact, the torque term $\bm{\tau}_{ij}$ should also be included \cite{ai2011}. 

As contact force model, we employed the linear spring-dashpot model based on a Hooke-type relation. 
In the linear spring-dashpot model, if the two spherical particles ($i$ and $j$) are in contact, namely their distance $r$ is less than their contact distance, $d$, ($d\equiv R_i+R_j$), the normal and tangential contact forces acting on $i$ are calculated respectively as \cite{Gu2014}:
\begin{equation}
	\bm{F}_{ij}^N=k_n \delta_{ij} \bm{n} _{ij} - \gamma_n m_{eff} \bm{v}_{n_{ij}} ,
	\label{EQ: Fn}
	\end{equation}
\begin{equation}
	\bm{F}_{ij}^T=k_t \delta_{ij} \bm{t} _{ij} - \gamma_t m_{eff}  \bm{v}_{t_{ij}} .
	\label{EQ: Ft}
	\end{equation}
In Eq. \ref{EQ: Fn} and \ref{EQ: Ft}, $k_n$ and $k_t$ are the normal and tangential elastic constants, $\gamma_n$ and $\gamma_t$ are viscous damping constants, $m_{eff}$ is the effective mass defined as $m_{eff}=\frac{m_i\cdot m_j}{m_i+m_j}$. Being $\mu$ the sliding friction coefficient, particle sliding occurs when the Coulomb criterion, namely $|\bm{F}_{t,ij}| < \mu|\bm{F}_{n,ij}|$, is not satisfied. 
As rolling friction model we used the Constant Directional Torque, CDT, in which the rolling friction is represented by a constant torque \cite{liggghts_doc, ai2011}.

The parameters we used in the simulations were evaluated as follow.
We firstly estimated a suitable value for the normal elastic constant, $k_n$, and then we assumed that:
\begin{equation}
	k_t=\frac{2}{7}k_n \,,\ \gamma_t=0 .
\end{equation}
The normal viscous dumping coefficient, $\gamma_n$, was set such that the restitution coefficient, $e_n$, which is defined as:
\begin{equation}
	e_n = \exp \left( \frac{-\gamma_n \pi}{\sqrt{\frac{4 k_n}{m_{eff}}-\gamma_n^2}} \right)
	\end{equation}
satisfy the desired value \cite{thornton2015, Chialvo2012, Gu2014}. 
Finally, the time step  was imposed smaller than the duration of collision \cite{Shafer1996}, whose estimation reads:
\begin{equation}
	t_n = \pi \left[ \frac{k_n}{m_{eff}}- \left( \frac{\gamma_n^2}{2} \right) \right]^{-1/2}.
	\end{equation}
The simulation parameters are summarized in Table \ref{TAB: parameter DEM} together with some system information.
\begin{table} \footnotesize
	\centering
	\begin{tabular}{llr} 
		\hline
		\textbf{Variable}            & \textbf{Symbol} & \textbf{Value}  	\\ \hline
		Particle diamater [m]		& $D_p$	  &  0.006 \\
		Intrinsic density of steel [kg/m$^3$]  & $\rho_S$        & 7800   \\
		Intrinsic density of glass [kg/m$^3$]  & $\rho_G$        & 2450    \\
		Normal elastic constant     	[N/m]			& $k_n$            & 8.00e05     \\ 
		Tangential elastic constant  [N/m]	 		& $k_t$            &    2.29e05 \\ 	
		Normal visco-elastic damping constant of steel [N/m] & $\gamma_{n,S}$ 	& 	9.61e03		\\
		Normal visco-elastic damping constant of glass [N/m] & $\gamma_{n,G}$ 	& 	1.71e04		\\
		Tangential visco-elastic damping constant [N/m] & $\gamma_t$ 		& 0 \\
		Restitution coefficient      & $e_n$           & 0.70    \\
		Inter-particle friction 		& $\mu$ 			  & 0.20		\\
		Rolling friction coefficient & $\mu _r$        & 0.005   \\
		Time step              & $\Delta t$      & $1e-6$  \\  \hline
		Wall-particle friction		& $\mu _{wp}$    & 0.33     \\
		Rotational speed [rpm]    & $\Omega$             & 23.44     \\ 
		Loading [N]    			& $\bm{F}_{load}$      & -10.72     \\ 
		Gravity [m/s$^2$]    		& $\bm{g}$      & -9.81    \\ 
		Boundary conditions          & $-$             & f f f      \\ \hline
		\end{tabular}
	\caption{A summary of the DEM simulation parameters.}
	\label{TAB: parameter DEM}
	\end{table}
The simulated annular shear cell had the same dimension of the real one (see Fig. \ref{FIG: geometry}) and finite boundary conditions in all directions. The top and bottom bumpy walls were represented with the same mesh used for the 3D printing of the real ones. On the top wall, $10.72$ N of loading acting on the negative \textit{z} direction was applied and the bottom wall was rotated at a constant \textcolor{black}{rotational speed} of 23.44 rpm. The sliding friction coefficient between particle and wall was set equal to 0.33. 

DEM simulations allowed the velocity profile of frictional and cohesionless particles made of steel and glass to be investigated. 

\clearpage
\section*{References}
\bibliography{bibfile}
\vspace{0.2cm}

\end{document}